\documentclass[prb,aps,twocolumn]{revtex4}
\usepackage{amsfonts,amsmath,amssymb,latexsym}
\usepackage{epsfig}

\begin{document}

\title{ RPAE versus RPA for the Tomonaga model with quadratic
energy dispersion}
\draft
\author {K. Sch\"onhammer}
\address{Institut f\"ur Theoretische Physik, Universit\"at
  G\"ottingen, Friedrich-Hund-Platz 1, D-37077 G\"ottingen}

\date{\today}

\begin{abstract}
Recently the damping of the collective charge (and spin) modes
of interacting fermions in one spatial dimension
was studied. It results
from the nonlinear correction to the energy dispersion in the
vicinity of the Fermi points. To investigate the damping one has to
replace the random phase approximation
 (RPA) bare bubble by a sum of more complicated diagrams.
It is shown here that a better starting point than the bare RPA
is to use the (conserving) linearized time dependent Hartree-Fock   
equations, i.e. to perform a
random phase approximation (with) exchange
 (RPAE) calculation. It is shown that 
the RPAE equation can be solved analytically for the special form of the 
two-body interaction often used in the Luttinger liquid framework. 
While (bare) RPA and RPAE agree for the case of a strictly linear
disperson there are qualitative differences for the case of the usual
nonrelativistic quadratic dispersion.

\end{abstract}

\maketitle
\vskip 2pc]
\vskip 0.1 truein

\section{Introduction}

In a seminal paper Tomonaga \cite{To} presented the exact
solution for the long wavelength 
density response of a system of {\it interacting}
fermions 
in one spatial dimension. In order to simplify the problem Tomonaga
studied the {\it high density limit} where the range of the
interaction is much larger than the interparticle distance.
 Then the Fourier transform $\tilde
v(k)$ of the two-body interaction is nonzero only for
values $|k|\le k_c$ where the cut-off $k_c$ is much smaller than the
Fermi momentum $k_c \ll k_F$. This implies that for not too strong
interaction the ground state and the low energy excited states   
have negligible admixtures of holes deep in the Fermi sea
and particles with momenta $|k|-k_F \gg k_c$. Therefore Tomonaga
{\it linearized } the nonrelativistic quadratic dispersion 
in the vicinity of the two Fermi points $\pm k_F$. He then
used Bloch's method of sound waves \cite{Bloch}
now called ``bosonization'' to obtain the ground state of the
interacting system as well as its low energy excitations.  
As an application Tomonaga calculated the
 density response of the
(spinless) system in the long wavelength limit. 
 This result was obtained  
before the pioneering paper by Bohm and Pines\cite{BP} in which the
authors  introduced
the ``random phase approximation'' (RPA) for the (linear) density
response of the (three-dimensional) interacting electron gas
in order to obtain e.g. the collective plasmon mode.
The RPA  is a good approximation in the high-density limit. Various 
alternative derivations of this approxmation can be given \cite{PN}, e.g.
using diagramatic techniques or linearizing the time dependent
Hartree-Fock (TDHF) equation \cite{GG} and neglecting the exchange contribution
and replacing the HF-eigenvalues by the noninteracting ones. 

If these additional approximations in the TDHF
equation are not made the corresponding
response functions are called RPAE, where the ``E'' stands for
exchange. While the RPA result is a simple expression in terms of
the response function of the noninteracting system, the calculation
of the RPAE response function usually requires the solution of
an integral equation.\cite{DBL,TL}   
 
For spin $1/2$  fermions collective modes usually occur in the charge 
and the spin density response. They were
studied by Dzyaloshinski and Larkin \cite{DL} using standard many
body techniques (avoiding bosonzation). They showed that
for linear dispersion  the
exact results for the collective modes for the Tomonaga
model are identical to the RPA result. Diagramatically this
can be understood in terms of the ``closed loop theorem'' 
which holds if one assumes strictly linear energy dispersion
around the two Fermi points \cite{DL,KHS}.

 Recently various activities 
started to study the charge and spin density response
when  deviations from a strictly  linear dispersion 
for one-dimensional fermions is taken into
account \cite{PK,PKKG,T1,Pereira}.
 Then the approximation using the ``bare bubble''
is no longer exact. 

It is one purpose of this paper to
point out that the bare RPA is not even a good starting
point for a perturbative calculation 
as e.g. it has the one-particle one-hole (1p-1h) 
continuum in the wrong place. It is much better
to use TDHF which is a conserving approximation \cite{Baym} as the starting
point to calculate the damping of the collective mode(s).  We show that 
it is possible to solve the RPAE equation for the Tomonaga model
analytically when the two-body interaction $\tilde v(q)$ takes
a constant value for $|q|<k_c$ and is zero otherwise. This exact
solution reduces to the RPA result with the noninteracting response
function (``bubble'') for strictly linear energy dispersion but
differs markedly for nonlinear dispersion. The continua are in
different places and the dispersion of the collective modes differ.
This is most drastically seen for the collective spin mode which
exists in RPAE but not in the bare RPA.
                 
\section{Linerarized time dependent Hartree-Fock approximation}

We want to describe the density response of a system of fermions on 
a ring of circumference $L$. The Hamiltonian is given by

\begin{equation}
H=\sum_k \epsilon_kc_k^\dagger c_k
+\frac{1}{2}\sum_{k_1,k_2,k_3,k_4}v_{k_1,k_2,k_3,k_4}
c_{k_1}^\dagger c_{k_2}^\dagger
c_{k_4}c_{k_3}~.
\end{equation}
 The index $k$ in addition to the momentum 
(possibly) includes a spin index. The first term presents the kinetic energy
with the nonrelativistic dispersion $\epsilon_k=k^2/2m$. The two-body
matrix element $v_{k_1,k_2,k_3,k_4} $  contains a factor
$\delta_{k_1+k_2,k_3+k_4}$ due to total momentum conservation. The 
remaining $k$-dependences are specified later.

In order to study the linear charge or spin density response the
Heisenberg equations of motion for $\langle c_k^\dagger
c_{k'}\rangle_t$ have to be solved in the presence of a
time dependent external 
one-body potential with matrix elements $V_{kk'}$. Using the
Hartree-Fock approximation to factorize the four fermion expectation
values and linearizing the expectation values around the HF ground state
\begin{equation}
\langle c_k^\dagger c_{k'}\rangle_t=\delta_{kk'}f(\epsilon_k^{\rm HF})
+\delta\langle c_k^\dagger c_{k'}\rangle_t
\end{equation}
yields
\begin{eqnarray}
\label{lTDHF}
\left[ i\frac{d}{dt}-(\epsilon_{k'}^{\rm HF}-\epsilon_{k}^{\rm HF})
\right ] 
\langle c_k^\dagger c_{k'}\rangle_t &=&\\
\left( f(\epsilon_k^{\rm HF})-f(\epsilon_{k'}^{\rm HF})\right )
[  V_{k'k}&+&\sum_{k_2,k_4}\bar v_{k'k_4k k_2} \delta 
\langle c_{k_4}^\dagger c_{k_2}\rangle_t ]~, \nonumber
\end{eqnarray}
where the HF-energies are given by
\begin{equation}
\label{HFev}
\epsilon_k^{\rm HF}=\epsilon_k+\sum_{k'}\bar v_{k,k',k,k'}
f(\epsilon_{k'}^{\rm HF}) 
\end{equation}
and $\bar v_{k_1,k_2,k_3,k_4}= v_{k_1,k_2,k_3,k_4}-v_{k_1,k_2,k_4,k_3}
$
are the antisymmetrized two-body matrix elements.
At $T=0$ the Fermi function is a step function 
not depending on the interaction $f(\epsilon_k^{\rm HF})=\Theta
(k_F-|k|)$.
 Eqs.
(\ref{lTDHF}) are the linearized TDHF equations for the expectation
values.\cite{GG}
 
As we are interested in the long wavelength response we put $k \to k_a$
and $k' \to k_a+q$, where $|q|\ll k_F$ and $k_a$ with $a=R,L$ is in the
neighborhood of the right or left Fermi point $\pm k_F$. 
Fourier transforming Eq. (\ref{lTDHF}) yields
\begin{eqnarray}
\label{RPAE}
&&\delta \langle c_{k_a}^\dagger c_{k_a+q}\rangle_\omega=
\frac{ f(\epsilon_{k_a}^{\rm HF})-f(\epsilon_{k_a+q}^{\rm HF}) }{\omega-
(\epsilon_{k_a+q}^{\rm HF}-\epsilon_{k_a}^{\rm HF} )} \\
&\times& \left[ V_{k_a+q, k_a}+
 \sum_{k'}\bar v_{k_a+q,k',k_a,k'+q}\delta \langle c_{k'}^\dagger
c_{k'+q} \rangle_\omega \right  ] ~,\nonumber 
\end{eqnarray}
where $\omega=\omega+i0$. With the assumption $|q|\ll k_F$ the 
$k'$ sum in Eq. (\ref{RPAE}) only can take values in the neighbourhood
of the two Fermi points. This implies that two different types of
two-body matrix elements contribute. For $k'$ in the neighborhood of
the same Fermi point as $k_a$
\begin{equation}
\label{g4}
\bar v_{k_a+q,\sigma;k'_a \sigma ';k_a ,\sigma ;k'_a+q,\sigma'}=
\frac{1}{L}\left [ \tilde g_4(q)-\delta_{\sigma\sigma'}\tilde
  g_4(k_a-k'_a) \right ]~.
\end{equation}

Here we have used the usual ``g-ology'' nomenclature \cite{So}
and explicitely  introduced the spin dependence. 
For the two-body matrix element with $k_a$ and $k'\approx k_{\bar a}$
where $\bar a $ differs from $a$ no exchange contribution 
occurs using Tomonaga's assumptions as this would involve a momentum
transfer of order $2k_F$, for which 
the two-body interaction is assumed to vanish. Thus
\begin{equation}
\label{g2}
\bar v_{k_a+q,\sigma;k'_{\bar a} \sigma ';k_a ,\sigma ;k'_{\bar a}+q,\sigma'}=
\frac{1}{L}~ \tilde g_2(q).
\end{equation}

We separate the two contributions in Eq. (\ref{RPAE}) and 
express the matrix elements of the external potential as 
$V_{k_a+q,\sigma,k_a,\sigma}=\tilde V_{\sigma,a}(q)/L$
\begin{eqnarray}
\label{RPAETo}
\delta \langle c_{k_a,\sigma}^\dagger c_{k_a+q,\sigma}\rangle_\omega=
\frac{1}{L}~\frac{ f(\epsilon_{k_a}^{\rm HF})-f(\epsilon_{k_a+q}^{\rm HF}) }{\omega-
(\epsilon_{k_a+q}^{\rm HF}-\epsilon_{k_a}^{\rm HF} )}~~~~~~~~~~~~~~~~~~~~\\
\times [\tilde V_{\sigma,a}(q)+
 \sum_{k',\sigma'}\tilde g_2(q) \delta \langle c_{k_{\bar a}',\sigma'}^\dagger
c_{k_{\bar a}'+q,\sigma'} \rangle_\omega ~~~~~~~~~~~~~~~\nonumber \\
 + \sum_{k',\sigma'}(\tilde g_4(q) -\delta_{\sigma \sigma'}
\tilde g_4(k_a-k'_a) ) \delta \langle c_{k_a',\sigma'}^\dagger
c_{k_a'+q,\sigma'} \rangle_\omega ]\nonumber ~.
\end{eqnarray}
For $\tilde g_2(q)= \tilde g_4(q)=\tilde v(q)$
this are the RPAE equations for the 
Galilei invariant Tomonaga model. Because of the 
exchange term in the $g_4$-interaction they usually have to be
solved numerically. \cite{PS} They simplify considerably 
for the special choice for the $\tilde g_m(q), (m=2,4)$
  mentioned in the introduction
\begin{equation}
\label{Kasten}
\tilde g_m(q)=\tilde g_m \Theta (k_c-|q|)~.
\end{equation}
With this assumption only expectation values of the type
\begin{equation}
\delta\langle \rho_{q,a,\sigma}\rangle_\omega =\sum_{k'_a} \delta \langle 
c_{k'_a,\sigma}^\dagger
c_{k'_a+q,\sigma} \rangle_\omega
\end{equation}
and $k_a' \to k_{\bar a}'$ occur on the rhs of Eq. (\ref{RPAETo}).
In Luttinger liquid terminology this are 
expectation values of the Fourier components of the densities 
of the right or left movers. 
 Therefore Eq. (\ref{RPAETo}), after
summing over $k_a$, yields a closed set of equations for the
$\delta\langle \rho_{q,a,\sigma}\rangle $. Dropping the index $\omega$
it reads 
\begin{eqnarray}
\label{mitspin}
\delta\langle
\rho_{q,a,\sigma}\rangle&=&R^a_{\rm HF}(q,\omega)[\tilde V_{\sigma,a}(q)
+\tilde g_4( \delta\langle
\rho_{q,a}\rangle  -\delta\langle
\rho_{q,a,\sigma}\rangle)\nonumber \\ 
 &&~~~~~~~~~~~~+\tilde g_2 \delta\langle
\rho_{q,\bar a}\rangle ]~,
\end{eqnarray}
with
\begin{equation}
\label{RHF}
R^a_{\rm HF}(q,\omega)=
\frac{1}{L}\sum_{k_a}
\frac{ f(\epsilon_{k_a}^{\rm HF})-f(\epsilon_{k_a+q}^{\rm HF}) }{\omega-
(\epsilon_{k_a+q}^{\rm HF}-\epsilon_{k_a}^{\rm HF} )}~,
\end{equation}
and the $\delta\langle
\rho_{q,a}\rangle $ are the charge densities for the right and left
movers
\begin{equation}
\delta\langle
\rho_{q,a}\rangle=\sum_\sigma \delta\langle
\rho_{q,a,\sigma}\rangle ~.
\end{equation}
With 
\begin{equation}
\label{Rtilde}
\tilde R^a(q,\omega)\equiv \frac{R^a_{\rm HF}(q,\omega) }{1+\tilde g_4
 R^a_{\rm HF}(q,\omega)}
 \end{equation}
Eq. (\ref{mitspin}) reads
\begin{eqnarray}
\label{general} 
\delta\langle
\rho_{q,a,\sigma}\rangle = \tilde R^a(q,\omega)
\left[ \tilde V_{\sigma,a}+\tilde g_4\delta\langle
\rho_{q,a}\rangle + \tilde g_2\delta\langle
\rho_{q,\bar a}\rangle \right ] 
\end{eqnarray}
This equation is the starting point for the calculation of the 
charge and spin response.

 The equations for the charge response are
obtained by summing Eq. (\ref{general}) over the spin index
\begin{eqnarray}
(1-2\tilde g_4 \tilde R^a)\delta\langle
\rho_{q,a}\rangle-2\tilde g_2 \tilde R^a\delta\langle
\rho_{q,\bar a}\rangle =2 \tilde R^a  \tilde  V_a^{(c)}
\end{eqnarray}
with $\tilde V_a^{(c)}\equiv \sum_\sigma \tilde V_{\sigma,a}/2$.
 The solution is given by
\begin{eqnarray}
\label{charge}
\delta\langle \rho_{q,a}\rangle_\omega
&=&\frac{2\tilde R^a(1-2\tilde g_4 \tilde R^{\bar a})\tilde V_a^{(c)}+4\tilde
  g_2  \tilde R^a  \tilde R^{\bar a}\tilde V_{\bar a}^{(c)} }
{(1-2\tilde g_4 \tilde R^a)(1-2\tilde g_4 \tilde R^{\bar a} )-4 \tilde
  g_2^2 \tilde R^a\tilde R^{\bar a}} \nonumber \\
&\equiv& R_{aa}(q,\omega) \tilde V_{a}^{(c)}
+R_{a\bar a}(q,\omega) \tilde V_{\bar a}^{(c)}
 \end{eqnarray}
The spin response $\delta \langle \sigma_{q,a} \rangle \equiv
\sum_\sigma \sigma\delta\langle
\rho_{q,a,\sigma}\rangle $ is obtained by taking the difference in
Eq. (\ref{general}). With $\tilde V_a^{(s)}\equiv \sum_\sigma \sigma 
\tilde V_{\sigma,a}/2$
one obtains
\begin{eqnarray}
\label{spin}
\delta \langle \sigma_{q,a} \rangle_\omega =2\tilde R^a~ \tilde V_a^{(s)}
=\frac{2R^a_{\rm HF}(q,\omega) }{1+\tilde g_4
 R^a_{\rm HF}(q,\omega)} ~\tilde V_a^{(s)}~.
\end{eqnarray}
For the spinless model originally studied by Tomonaga the two 
$\tilde g_4$ terms on the rhs of Eq. (\ref{mitspin}) cancel and 
one obtains \cite{PP}
\begin{equation}
\label{spinless}
\delta \langle \rho_{q,a}\rangle_\omega=\frac{R^a_{\rm HF}\tilde V_a+
\tilde g_2R^{\bar
    a}_{\rm HF}R^a_{\rm HF} \tilde V_{\bar a}}
{1-\tilde g_2^2 R^{\bar
    a}_{\rm HF}R^a_{\rm HF}}~.
\end{equation}
Equations  (\ref{charge}), (\ref{spin}),
and (\ref{spinless}) are the central results of this
paper. The general behaviour and limiting cases 
of the corresponding response functions are discussed in the
following section.

\section{General results and limiting cases}

In this section we discuss the RPAE charge and spin response
restricting ourselves to zero temperature. The basic building blocks of
the RPAE response functions are the functions $\tilde R^a(q,\omega)$
which via Eq. (\ref{Rtilde}) are expressed in terms of the
$R^a_{\rm HF}(q,\omega)$. In order to calculate these functions we first
have to determine the HF-eigenvalues $\epsilon^{\rm HF}_k$ defined in
Eq. (\ref{HFev}). For the special choice of the two-body matrix
elements in Eqs. (\ref{g4}), (\ref{g2}), and (\ref{Kasten}) one obtains
for $k=\pm k_F+\tilde k$ with $|\tilde k|<k_c$ and $k_c \le k_F/2$
\begin{eqnarray}
 \epsilon^{\rm HF}_{\pm k_F+\tilde k}&=& \frac{(\pm k_F+\tilde k )^2}{2m}
\pm \frac{\tilde g_4}{2\pi}\tilde k +\mbox {const.}\\ 
&=& \pm (v_F+\frac{\tilde g_4}{2\pi})\tilde k+\frac{\tilde
  k^2}{2m}+\tilde c \nonumber  
\equiv \pm v_F^{\rm HF}\tilde k +\frac{\tilde
  k^2}{2m}+\tilde c ~.
\end{eqnarray}
 The constant is given by the Hartree contribution and the constant 
part of the Fock term. It is not needed here, as the Fermi function at $T=0$ is
independent of the interaction and and therefore
only differences of HF-energies occur in Eq. (\ref{RHF}) .   

For $|q|<k_c$ the behaviour of the response functions can be 
expressed in terms  of two types of
dimensionless quantities using  $m=k_F/v_F$.
\begin{equation}
\alpha_i\equiv \frac{\tilde g_i}{2\pi v_F}~,~~~~~
\beta \equiv \frac{q}{2k_F}~.
\end{equation} 
If we define $s_a=1$ for $a=R$ and $s_a=-1$ for $a=L$ 
the HF response functions
in Eq. (\ref{RHF}) are given by
\begin{equation}
\label{HFR}
R^a_{\rm HF}(q,\omega)=-\frac{1}{2\beta}\frac{1}{2\pi v_F}\log \left
[ \frac{1+\alpha_4+\beta-s_a\tilde \omega}{1+\alpha_4-\beta-s_a\tilde \omega}
\right ]~,
\end{equation} 
with $\tilde \omega\equiv \omega/(v_Fq)$. 

The linear dispersion limit
corresponds
to the high density limit $\beta \to 0$ ($k_F \to \infty$)
\begin{equation}
\beta \to 0 ~:~~~~~R^a_{\rm HF}(q,\omega)\to \frac{s_a}{2\pi v_F}
\frac{1}{\tilde \omega -s_a(1+\alpha_4)}
\end{equation} 
In this limit the functions  $\tilde R^a$ reduce to the noninteracting
ones for linear dispersion
\begin{equation}
\tilde R^a(q,\omega) \to \frac{q}{2\pi}\frac{s_a}{\omega -s_av_Fq}\equiv 
R^a_0(q,\omega)_{\rm lin}~~.
\end{equation} 
The well known results for the charge and spin mode of the Tomonaga
model for linear dispersion \cite{DL,Giamarchi,KS} then follow from 
Eqs. (\ref{charge}) and 
 (\ref{spin}) as
\begin{eqnarray}
\label{linearmodes}
\omega_{q,c}^2&=& (v_Fq)^2\left [ (1+2\alpha_4)^2-(2\alpha_2)^2\right]
\nonumber \\
\omega_{q,s}^2&=& (v_Fq)^2~.
\end{eqnarray}
For the spinless model Eq. (\ref{spinless}) yields the same form
for the charge mode as in
Eq. (\ref{linearmodes}) but with $2\alpha_i \to \alpha_i$.  

\subsection{The $g_4$ model}

While the spin response function in Eq. (\ref{spin}) only depends
on  the
$g_4$ interaction,  the charge response functions $R_{aa'}$ in
Eq. (\ref{charge}) involve a coupling of
both Fermi points via the $g_2$ interaction .
 The chiral model obtained by putting $\tilde g_2=0$
is called the $g_4$ model \cite{Giamarchi,KS}. We discuss its 
 response functions for the right movers and  $q>0$. The
charge response follows from Eq. (\ref{charge}) as
\begin{eqnarray} 
\label{g4charge}
\delta \langle \rho_{q,R}\rangle_\omega&=& \frac{ 2 \tilde
  R^R(q,\omega) }{1-2\tilde g_4\tilde
  R^R(q,\omega)}\tilde V_R^{(c)}(q,\omega) \nonumber \\
(\beta \to 0) &\to& \frac{2R^R_0(q,\omega)_{\rm lin} }{1-2\tilde g_4
  R^R_0(q,\omega)_{\rm lin}}\tilde V_R^{(c)}(q,\omega)~.
\end{eqnarray}
In the second line we also presented the result for  the linear
dispersion case. It looks 
tempting to generalize it to nonlinear dispersion by simply
dropping the index ``lin'' and use instead 
the noninteracting response function for quadratic dispersion
which is obtained by putting $\alpha_4=0$ in Eq. (\ref{HFR}).
As pointed out in the introduction this would be a bad
approximation as it e.g. has the nonvanishing imaginary part of the
response function in the wrong frequency range. To show this
we use Eq. (\ref{Rtilde})
to express $\tilde R^R$ in terms of $R^R_{\rm HF}$ 
\begin{eqnarray}
\label{g4charge2} 
\delta \langle \rho_{q,R}\rangle_\omega= \frac{2
  R^R_{\rm HF}(q,\omega)}   
{1-\tilde g_4
  R^R_{\rm HF}(q,\omega)}~\tilde V_R^{(c)}(q,\omega) 
\end{eqnarray}
For $q>0$  the continuous part of the imaginary part
 of this RPAE response
function corresponding to the HF 1p-1h
excitations
 is different from zero for
$v_Fq(1+\alpha_4-\beta)\le \omega \le v_Fq(1+\alpha_4+\beta)$, while the
 imaginary part of $R_0^R(q,\omega)$ is different from zero for
$v_Fq(1-\beta)\le \omega \le v_Fq(1+\beta)$.

The RPAE charge mode follows from Eq. (\ref{g4charge2}) by putting
the denominator to zero. The solution of the resulting transcendental
equation is given by  
\begin{equation}
\omega^{\rm RPAE}_{q,c}
=v_Fq \left (1+\alpha_4+\beta \coth{\frac{\beta}{\alpha_4}}\right )~,
\end{equation}
which in the limit $\beta \to 0$ agrees with the $\alpha_2=0$
result in Eq. (\ref{linearmodes}). The mode is always above the
the HF 1p-1h continuum  as also shown in Fig. 1. This is not the
case for the RPA-charge mode which is obtained from $1-2\tilde g_4
R_0^R=0$
as
\begin{equation}
\omega^{\rm RPA}_{q,c}
=v_Fq \left (1+\beta \coth{\frac{\beta}{2\alpha_4}}\right )~.
\end{equation}
It follows from Eq. (\ref{spin}) that the RPAE collective spin mode is
generally independent of $\tilde g_2$. Its location is given by
\begin{equation}
\omega^{\rm RPAE}_{q,s}
=v_Fq \left (1+\alpha_4-\beta \coth{\frac{\beta}{\alpha_4}}\right )~,
\end{equation}  
which in the limit $\beta \to 0$ agrees with the 
result in Eq. (\ref{linearmodes}) which is independent of the 
interaction strength for the Tomonaga model with linear dispersion.
 The RPA spin response function
$2R^R_0(q,\omega)$ has only a continuous non-vanishing imaginary
part.
This is different for the RPAE result for nonlinear dispersion, where 
 the delta spike of the spin mode is always below the HF 1p-1h
continuum.  

In Fig. 1 we show the dispersion of the modes relative to the
HF-continuum for $\alpha_4=0.5$.

\vspace{0.0cm}    
\begin{figure}[tb]
\begin{center}
\vspace{-0.0cm}
\leavevmode
\epsfxsize7.5cm
\epsffile{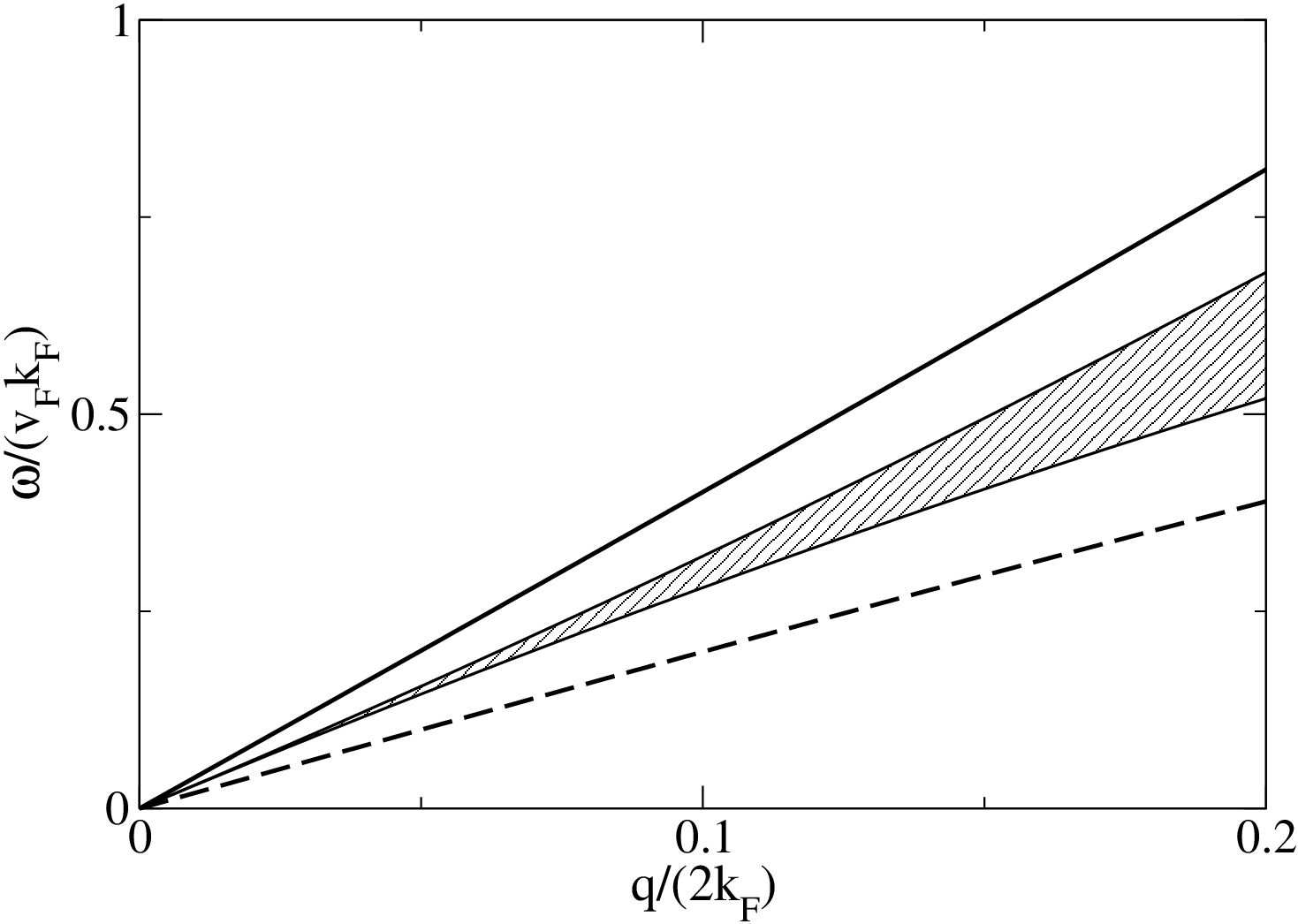}
\caption {Dispersion $\omega^{\rm RPAE}_{q,c(s)}/(v_Fk_F)$
of the collective charge and spin modes for the
$g_4$-model including spin for a repulsive interaction of dimensionless
strength $\alpha_4=\tilde g_4/(2\pi v_F)=0.5 $. The charge mode (full
line) lies above the HF 1p-1h continuum and the spin mode (dashed
line)
below it. This continuum is indicated by the hatched area.}   
\end{center}
\vspace{0.5cm}
\end{figure}

 The imaginary part of the
RPAE spin response function as a function of 
$\omega$ for fixed $\beta=0.1$ is shown in Fig. 2, where it is 
compared to the result of the (bare) RPA result, which
shows no collective mode. 
The weight of the RPAE collective spin mode which is slightly below $\tilde
\omega =1$ carries 94.8 per cent  of the total RPAE spectral weight.
\vspace{0.5cm}    
\begin{figure}[tb]
\begin{center}
\vspace{-0.0cm}
\leavevmode
\epsfxsize7.5cm
\epsffile{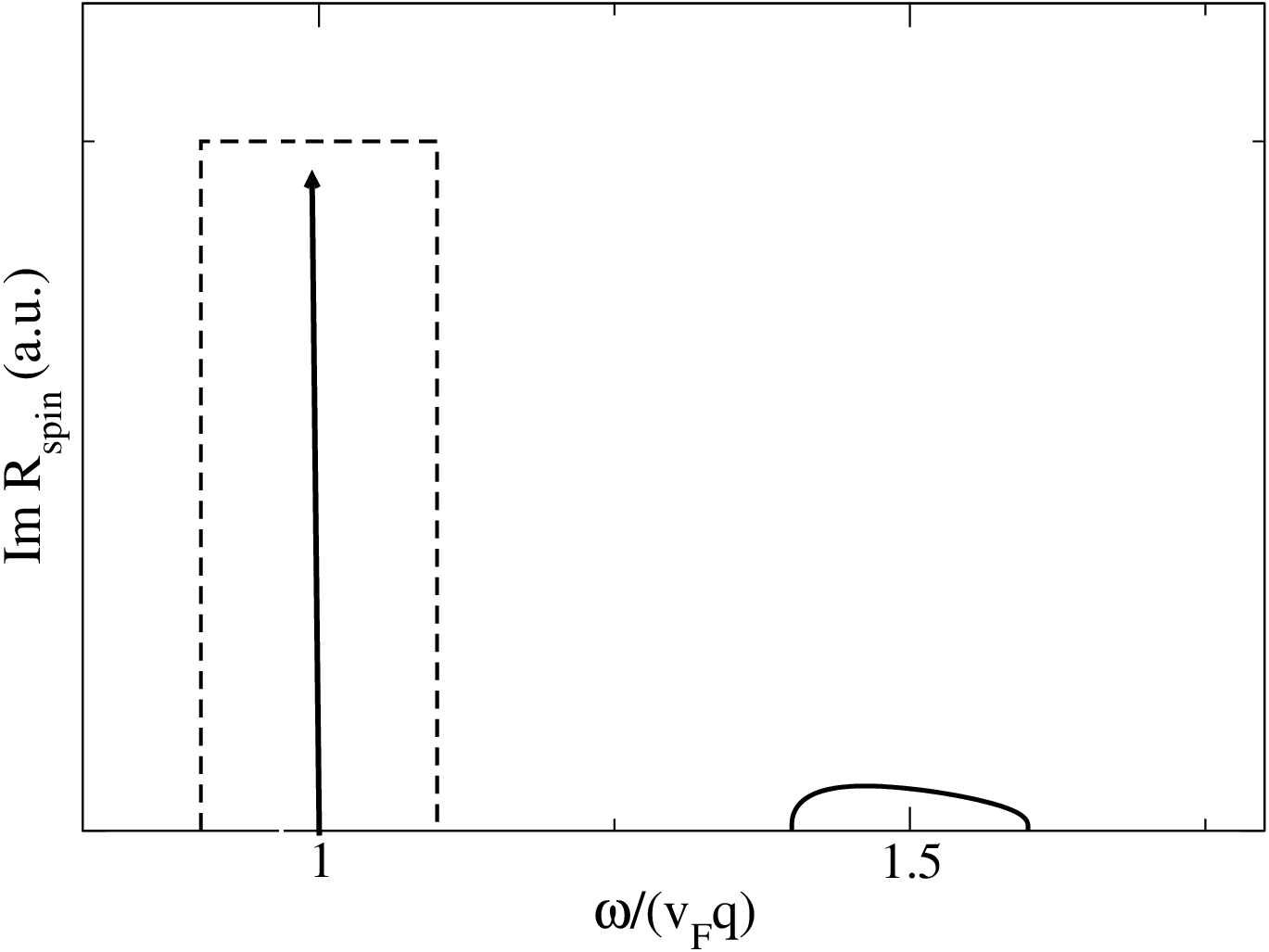}
\caption {Imaginary part of the spin response function for the
$g_4$ model for $q/k_F=0.2$
 as a function of $\omega$ for $\alpha_4=0.5$, as in Fig. 1.
The full line shows the RPAE result, with the collective spin
mode indicated by the arrow. The dotted curve shows the RPA result
which only has a continous part but no collective spin mode.    }
\label{Levitov01}
\end{center}
\end{figure}

\subsection{Galilei invariant model:  $\tilde g_2=\tilde g_4=\tilde v_0$}

In this subsection we discuss the model that describes interacting 
fermions with a spin independent 
two-body interaction $v(|x_i-x_j|)$ on a ring with

\begin{equation}
 \tilde v(q)=\tilde v_0\Theta(k_c-|q|)~,
\end{equation}
and $k_c\le k_F/2$.

Then the expression for the total charge response in
Eq. (\ref{charge})
simplifies to
\begin{eqnarray}
\label{Galileicharge}
R(q,\omega)\equiv \sum_{a,a'}R_{aa'}(q,\omega)&=&
\frac{\tilde R(q,\omega)}{1-\tilde v_0 \tilde R(q,\omega)} \\
(\beta \to 0) &\to&
\frac{ R_0(q,\omega)_{\rm lin}}{1-\tilde v_0 
 R_0(q,\omega)_{\rm lin}} \nonumber
\end{eqnarray}
where in $\tilde R\equiv 2(\tilde R^R+\tilde R^L)$ and
 $ R_0\equiv 2( R_0^R+ R_0^L)$
the factor two is the due to the spin degeneracy. For the spinless
 model one obtains the same result but this factor of two is missing.

 As in the previous subsection
it would be a bad approximation to generalize the bare RPA in
the second line in Eq. (\ref{Galileicharge})
(which provides the exact result for linear
dispersion) by simply dropping the index ``lin''. 

The transcendental equation which results from putting the
denominator in Eq. (\ref{Galileicharge}) equal to zero can no longer
be solved analytically. In Fig. 3 we show numerical results for
$v_c(q)\equiv \omega_{q,c}/|q|$ for 
the spinfull model for $\alpha=\tilde v_0/(2\pi
v_F)=0.25$.

\begin{figure}[tb]
\begin{center}
\leavevmode
\epsfxsize7.5cm
\epsffile{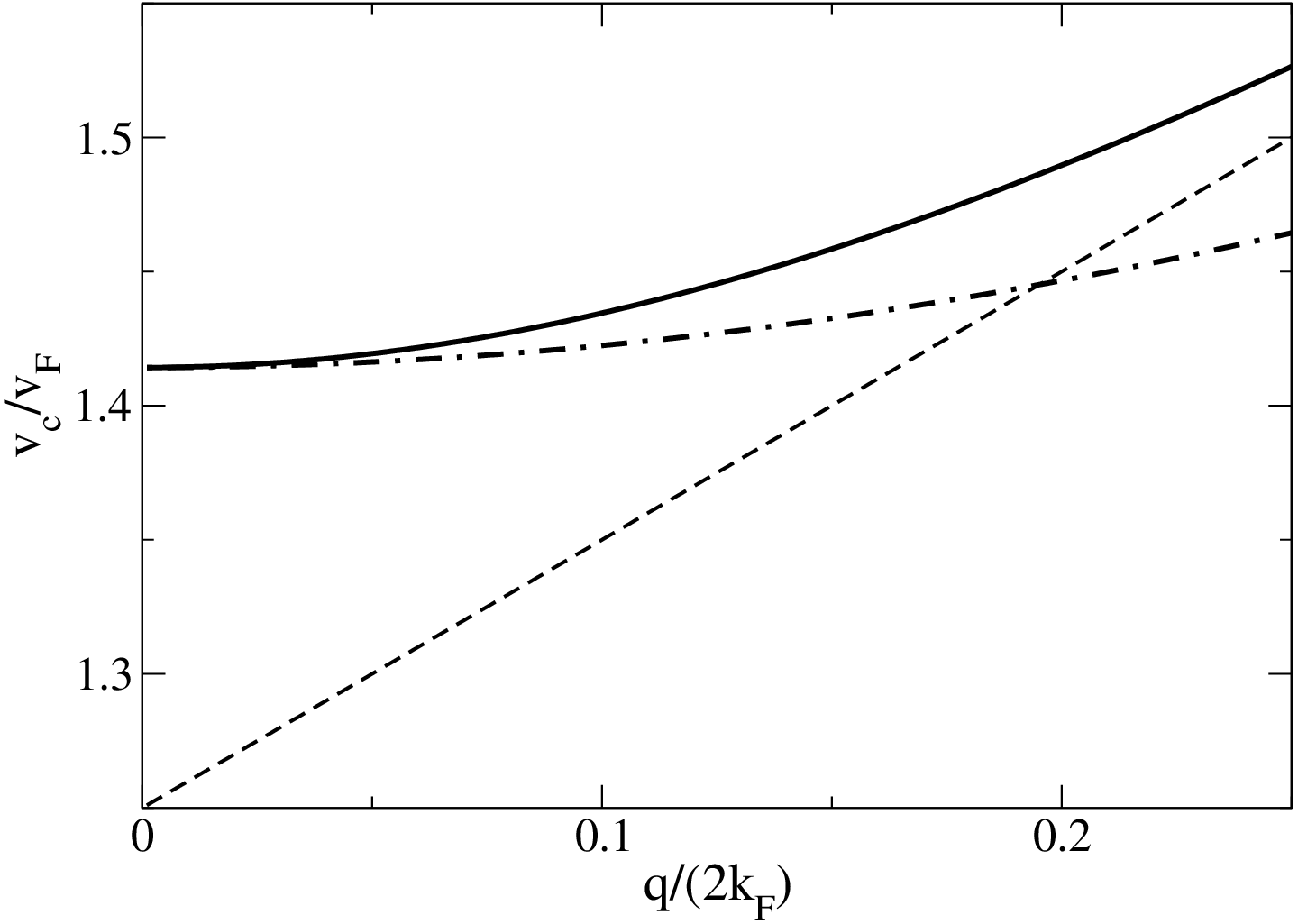}
\caption {  Dispersion of the normalized charge velocity $v_c(q)/v_F
\equiv \omega_{q,c}/(qv_F)$
 for the Galilei invariant
model ($\tilde g_2$=$\tilde g_4$)
 including spin for a repulsive interaction of dimensionless
strength $\alpha=\tilde v_0/(2\pi v_F)=0.25 $. The RPAE result (full
line) lies above $1+\alpha+\beta$  indicated by the
dashed line, marking the top of the HF 1p-1h continuum.
 The RPA charge velocity (dashed-dotted line) crosses
the dashed line
 and only stays above $1+\beta$ 
 corresponding to the bare 1p-1h continuum (not shown).  }
\end{center}
\end{figure}

 As a comparison we also show the corresponding bare
RPA result. For $0<\alpha<2$ the charge velocity for linear dispersion
$v_c=v_F\sqrt{1+4\alpha}$ is larger than $v_F^{\rm HF}=v_F(1+\alpha)$.
 For these 
$\alpha$ values the RPAE charge mode is above the HF
1p-1h continuum.
For larger repulsive interactions $\alpha>2$ the  charge velocity for
 linear dispersion is smaller than $v_F^{\rm HF} $ which puts the RPAE 
charge mode below the HF 1p-1h continuum. 

 In the spinless model the RPAE charge mode  for all
$\alpha >0$ lies below the HF 1p-1h continuum. This can easily be seen
by examining the denominator in Eq. (\ref{spinless}). This is 
qualitatively different from the RPA result which has the charge
mode above the noninteracting continuum.

In an approximation beyond RPAE the collective modes
discussed above are expected
to aquire a finite lifetime leading to a broadening of the delta peaks
in the imaginary part of the response function.  
Because the higher HF p-h continua have an additional extent
for energies above 
the 1p-1h continuum but not below it,  the damping of the collective
modes will depend crucially on how they are located relative to the 
HF  1p-1h continuum.

\section{Summary}

In this paper we have revisited the old problem of the
long wavelength linear
charge and spin response of interacting fermions in one spatial
dimension. As there is renewed interest in the questions of
the lifetime of the ideal bosonic modes of the Tomonaga model
due to deviations from the linear electronic dispersion\cite{PK,PKKG,T1} we 
hope to have clarified the often quoted statement that ``RPA becomes
exact in the long wavelength limit''. We have shown 
that a naive generalizaton of this result fails badly.  

We have not presented a calculation for the lifetimes but only
set the proper stage from where to start a perturbative
approach. In order to avoid tadpole diagrams HF-propagators 
have to be used. The use of bare propagators is useful only
for the case of linear dispersion because of the ``closed loop
theorem''.\cite{DL,KHS}

 For the spinfull model and not too strong repulsive
interactions the collective charge and spin modes are above
respectively below the HF 1p-1h continuum.
As the higher particle-hole continua for fixed $q$
with particle hole pairs around both Fermi points
are higher in energy the
damping of the spin mode is expected to be quite different from
the charge mode. For the spinless model and repulsive 
interaction the RPAE charge mode is always below the 1p$-$1h
 continuum. Therefore
we also expect the spinfull and the spinless model to behave
differently 
concerning the damping of the charge mode for weak repulsive
interactions.

\section{Acknowledgements}

The author is grateful to P. Kopietz and V. Meden for valuable
discussions.

\end{document}